\documentclass[preprint,12pt]{elsarticle}
\usepackage{graphicx}

\usepackage{amssymb}
\journal{Journal of Magnetism and Magnetic Material}
\begin{document}
\begin{frontmatter}

\title{Magnetic hyperfine field at \textit{s-p} impurities on Laves phase compounds}

\author[aff1]{C. M. Chaves}
\ead{cmch@cbpf.br}
\author[aff2]{A. L. de Oliveira} 
\author[aff3]{N. A. de Oliveira} 
\author[aff1]{A.Troper} 
\address[aff1]{Centro Brasileiro de Pesquisas F\'\i sicas, Rua Xavier Sigaud 150, Rio de Janeiro, 22290-180, Brazil}
\address[aff2]{ Instituto Federal de Educa\c{c}\~{a}o, Ci\^{e}ncia e Tecnologia do Rio de Janeiro, Campus Nil\'{o}polis, Rua L\'{u}cio Tavares 1045, 26530-060, Nil\'{o}polis, RJ, Brazil}
\address[aff3]{Universidade do Estado do Rio de Janeiro, Rua S. Francisco Xavier 524,  Rio de Janeiro, 20550-013, Brazil}

\begin{abstract}

Recent experimental results for the magnetic hyperfine field $B_{\rm hf}$ at the nuclei of \textit{s-p} impurities such as  ${}^{119}$Sn in intermetallic Laves phases  $RM_2$ ($R$ = Gd, Tb, Dy, Ho, Er; $M$ = Fe, Co) and ${}^{111}$Cd in $R$Co$_{2}$, the impurity occupying a $R$ site indicate that the ratio $B_{\rm hf}/\mu_{3d}$ exhibits different behavior when one goes from $R$Fe$_2$ to $R$Co$_2$.  In this work, we calculate these local moments and the magnetic hyperfine fields.  In our model, $B_{\rm hf}$ has two contributions: one arising from the $R$ ions, and the other arising from magnetic $3d$-elements; these separate contributions allow the identification of the origin of different behavior of the  ratio  mentioned above. For ${}^{111}$Cd in $R$Co$_{2}$ we present also the contributions for $B_{\rm hf}$ in the light rare earth Pr, Nd, Pm, Sm compounds. For the sake of comparison we apply also the model to ${}^{111}$Cd diluted in $R$Ni$_2$.  Our self-consistent magnetic hyperfine field results are in good agreement with those recent experimental data.

\end{abstract}

\end{frontmatter}

\section{The model}

The local magnetic moment at a Sn impurity diluted in $R$Fe$_{2}$ and $R$Co$_{2}$ or Cd in $R$Co$_{2}$, 
has the following contributions: one from the $R$ ions, consisting of a polarization of the $\textit{s-p}$ impurity level by 4f and 5d rare-earth electrons and from a potential ($V$) due to the presence of the impurity. Also included in $V$, is the nearest-neighbor hopping change due to the break of translational invariance by the impurity. The other one, from the transition metal, is a magnetic field produced by the neighbors of the impurity, which occupies a rare earth site (see Ref.( \cite {Oliveira2003}) and references therein). 

The Hamiltonian to describe the formation of the local {\it s-p} magnetic moment and hyperfine field is
\begin{equation}
\mathcal{H}=H_{R}+V+H_{M}.\label{eq:ham1}
\end{equation}
In Eq.~(\ref{eq:ham1}),
\begin{equation}
H_{R} = \sum_{ i,\sigma }\varepsilon _{\sigma }^{{\rm h}}c_{i\sigma}^{\dagger}c_{i\sigma }
+\sum_{i,j, \sigma }t_{ij}c_{i\sigma }^{\dagger}c_{j\sigma},
\label{eq:HR}
\end{equation}
defines an effective pure rare earth host which consists in a conduction \textit{s-p} band polarized by the 4f and 5d electrons.  In Eq.~(\ref{eq:HR}), $\varepsilon _{\sigma }^{{\rm h}}$ is the center of the {\it s-p} energy band, now depending on the spin $\sigma $ orientation, $c_{i\sigma}^{\dagger}$ ($c_{i\sigma}$) is the creation (annihilation) operator of conduction electrons at site $i$ with spin $\sigma $ and $t_{ij}$ is the electron hopping energy between neighboring $i$ and $j$ sites.
The second term of Eq.~(\ref{eq:ham1}) is the potential due to the presence of the impurity at site $i=0$.
\begin{equation}
V=\sum_{\sigma}{V_{0\sigma }c_{0\sigma }^{\dagger}c_{0\sigma } }
+\tau\sum_{l\neq 0,\sigma }t_{0l}\left( c_{0\sigma }^{\dagger }c_{l\sigma }+c_{l\sigma }^{\dagger }c_{0\sigma }\right), 
\label{eq:V}
\end{equation}
where $V_{0\sigma}=(\varepsilon _{0\sigma }^{\rm I}-\varepsilon _{\sigma }^{\rm h})$  is a spin dependent local term, $\varepsilon _{0\sigma }^{\rm I}$ being the {\it s-p} impurity state energy level. The parameter $\tau $ takes into
account the change in the hopping energy associated with the
presence of the impurity~\cite{Oliveira2003,Acker91b,Oliveira95},$\tau=0$ meaning no disorder in the hopping. 

The last term of Eq.~(\ref{eq:ham1}),
\begin{equation}
H_{M}=-\sum_{l\neq 0,\sigma }\sigma J^{{\rm sd}}\left\langle S^{M}\right\rangle c_{l\sigma }^{\dagger }c_{l\sigma },
\end{equation}
is the interaction energy between the magnetic field from the $M$ ions and the impurity spin. $J^{{\rm sd}}$ is an exchange parameter  and $\left\langle S^{M}\right\rangle $ is the average magnetic moment at $M$ sites.

Let us first solve the problem with $H_{M}=0$. 
Using Dyson equation, the
exact Green's functions $\widetilde{g}_{ij\sigma }(z)$ due to the charge perturbation 
at the origin, is~\cite{Oliveira2003}
\begin{eqnarray}
\widetilde{g}_{ij\sigma }(z) &=& g_{ij\sigma }(z)+g_{i0\sigma }(z)
\frac{V_{{\rm eff}}^{\sigma }(z)}{\alpha^{2} - g_{00\sigma }(z)V_{{\rm eff}}^{\sigma }(z)}g_{0j\sigma }(z)  \nonumber \\
&&+\left( \alpha - 1\right)^{2} \frac{
g_{00\sigma }(z)\delta _{i0}\delta _{0j}}{\alpha ^{2}-g_{00\sigma }(z)V_{{\rm eff}}^{\sigma }(z)}   \nonumber \\
&&-\left( \alpha - 1\right) \frac{\alpha \left( g_{i0\sigma }(z)\delta _{0j}+\delta
_{i0}g_{0j\sigma }(z)\right) }{\alpha ^{2}-g_{00\sigma }(z)V_{{\rm eff}}^{\sigma }(z)}.
\label{eq:4}
\end{eqnarray}
Here $g_{ij\sigma }(z)$ is the Green's functions for the pure host with $z=\varepsilon+i0$, and $\alpha = \tau +1$. 
The effective potential $V_{{\rm eff\,}}^{\sigma }(z)$ is given by: 
\begin{equation}
V_{{\rm eff}}^{\sigma }(z)=V_{0\sigma }+(\alpha ^{2}-1)(z-\varepsilon^{\rm h}_{\sigma } ).,
\end{equation}

For the full Hamiltonian (\ref{eq:ham1}), the perturbed
Green's functions $G_{ij\sigma }(z)$ is 
\begin{equation}
G_{ij\sigma }(z)=\widetilde{g}_{ij\sigma }(z)
+\sum_{l\neq 0}\widetilde{g}_{il\sigma }(z)T_{ll}^{\sigma }\widetilde{g}_{lj\sigma }(z)
\end{equation}
with: 
\begin{equation}
T_{ll}^{\sigma }=\frac{V_{nn}^{\sigma }}{1-g_{ll\sigma }(z)V_{nn}^{\sigma }},
\end{equation}
where $V_{nn}^{\sigma } = -\sigma Z_{\rm nn}J^{sd}\left\langle S^{M}\right\rangle$ is the magnetic coupling between the impurity and the $M$ sites and $Z_{{\rm nn}}$ is the number of the nearest neighbor $M$ ions
surrounding a \textit{s-p} (Sn or Cd) impurity. 
At this point we adopt the following approximation: one consider only the contribution of nearest neighbor $M$ sites; and in this sense we have a cluster-like approach to calculate the contribution to the magnetic moment arising from $M$ ions. 
We assume that $V_{\rm nn}^{\sigma }$ is small 
compared to  $V_{0\sigma }$ thus justifying the Born approximation, $T_{ll}^{\sigma }\approx V_{\rm nn}^{\sigma }$.
The local Green's functions $G_{00\sigma }(z)$ at the
impurity site is then 
\begin{eqnarray}
G_{00\sigma }(z)&=&\frac{g_{00\sigma }(z)}{\alpha ^{2}-g_{00\sigma }(z)\,V_{\rm eff}^{\sigma }(z)}
+ 
\frac{\alpha ^{2}V_{{\rm nn}}^{\sigma
}}{\left[ \alpha ^{2}-g_{00\sigma }(z)\,V_{\rm eff}^{\sigma }(z)\right] ^{2}} \nonumber \\
&&\times \left[
\frac{\partial g_{00\sigma }(z)}{\partial z}+\left(
g_{00\sigma }(z)\right) ^{2}\right] . 
\end{eqnarray}
Note that in the local Green's functions $G_{00\sigma }(z)$, the
first term is due to the rare-earth ions whereas the second comes from
the polarization produced by the $M$ ions.  The local potential $V_{0\sigma }$
is self consistently determined using the Friedel screening condition (see for instance Ref.~\cite
{Oliveira95}) for the total charge difference $\Delta Z$ between impurity and
rare earth atoms.
\begin{figure}[htbp]
\includegraphics[angle=0,width=0.60\textwidth]{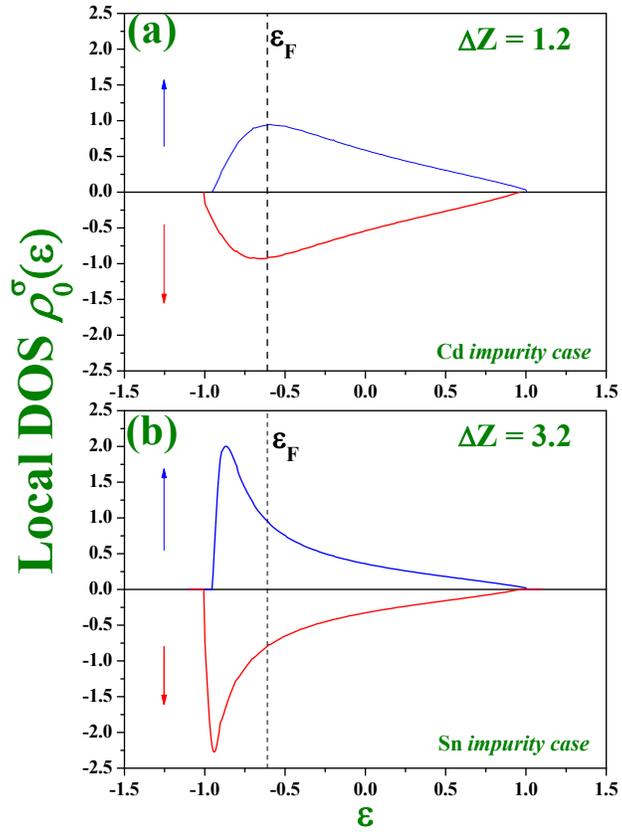}
\caption{Calculated local density of states $\rho ^{\sigma }_{0}(\varepsilon)$ (a) at the Cd and (b) at the Sn impurity site both in GdCo$_{2}$. Due to the splitting and the band deformation caused by the scattering produced by the charge difference between the impurity and host, these local DOS's will generate local magnetic moments in the indicated directions (see text). } \label{fig:dos}
\end{figure} 

Assuming that the screening of this charge
difference is made by the {\it s-p} band then $\Delta Z=\Delta Z_{\uparrow }+\Delta Z_{\downarrow },$ where $
\Delta Z_{\sigma }$ is 
\begin{equation}
\Delta Z_{\sigma }=-\frac{1}{\pi }{\rm Im}\ln \left[ \alpha ^{2}-g_{00\sigma }(\varepsilon_{\rm F})\,V_{\rm eff}^{\sigma
}(\varepsilon_{\rm F})\right] ,
\end{equation}
where $\varepsilon_{\rm F}$ is the Fermi energy level. The local {\it s-p} density of states per spin direction at
the impurity site are calculated by $\rho _{\sigma }(\varepsilon
)=\left( -1/\pi \right) {\rm Im\,}G_{00\sigma }(z)$. \ The
local {\it s-p} electron occupation number, $n_{0\sigma
}$, is obtained by integrating the corresponding
local density of states up to the Fermi level $\varepsilon _{{\rm F}}$. \ The
total magnetic moment ($\widetilde{m_{0}}$) at a {\it s-p} impurity, given by $\widetilde{m_{0}}=n_{0\uparrow }-n_{0\downarrow }$ is 
\begin{equation}
\widetilde{m_0}=\widetilde{m}_{0}^{R}+\widetilde{m}^{{\rm ind}}_{0},
\end{equation}
where 
\begin{equation}
\widetilde{m}_{0}^{R}=-\frac{1}{\pi }\sum_{\sigma }\int_{-\infty
}^{\epsilon _{{\rm F}}}{\rm Im\,}\,\frac{\sigma\; g_{00\sigma }(z)}{\alpha ^{2}-g_{00\sigma }(z)V_{\rm eff}^{\sigma
}(z)}\,{\rm d}z  \label{eq:momR}
\end{equation}
is the contribution from rare-earth ions and 
\begin{eqnarray}
\widetilde{m}^{{\rm ind}}_0&=&+\frac{1}{\pi }\sum_{\sigma
}\int_{-\infty }^{\epsilon _{{\rm F}}}{\rm Im\,}\frac{\alpha ^{2}Z_{\rm nn}J^{{\rm sd}}\left\langle S^{M}\right\rangle }{\left[ \alpha
^{2}-g_{00\sigma }(z)\;V_{\rm eff}^{\sigma
}(z)\right] ^{2}} \nonumber \\ &&\times \left[ \frac{\partial g_{00\sigma 
}(z)}{\partial z}+\left( g_{00\sigma }(z)\right) ^{2}\right] \;{\rm d}z
\end{eqnarray}
is the contribution from the $M$ nearest neighbor ions.  The total magnetic
hyperfine field at the impurity site is 
\begin{equation}
B_{hf}=A(Z_{{\rm imp}})\widetilde{m}_{0},
\label{chf}
\end{equation}
where $A(Z_{{\rm imp}})$ is the Fermi-Segr\`{e} contact coupling parameter. Similarly we define $B_\mathrm{hf}^\mathrm
{ind}$ and $B_\mathrm{hf}^{R}$.

\section{Results}
\begin{figure}[htbp]
\includegraphics[angle=0,width=0.60\textwidth]{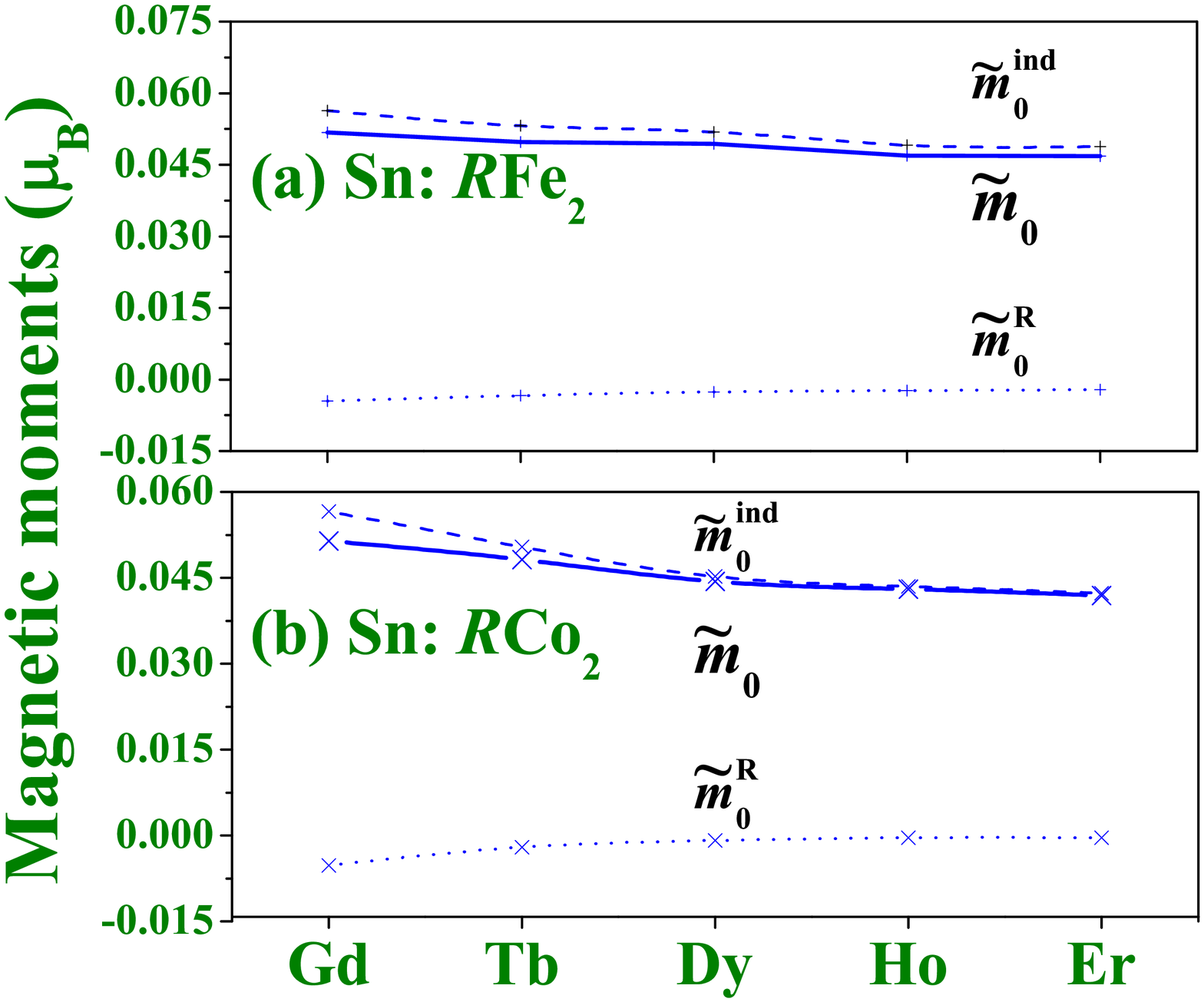}
\caption{Calculated total magnetic moment $\widetilde{m_0}$ at Sn impurity diluted in (a) $RFe_{2}$ and (b) $RCo_2$ (solid lines).  The dotted lines correspond to the contribution from the rare earth and the dashed lines correspond to the contribution of the $M$ ions.} \label{fig:mmSn}
\end{figure}
Our calculations are restricted to compounds where the rare earth ions are trivalent. This precludes Ce$M_{2}$, Yb$M_{2}$and Eu$M_{2}$ compounds: Ce is in a mixed valence state and Yb and Eu are divalent.
In order to calculate the local moments and the magnetic hyperfine fields at a Sn or Cd impurity diluted in $RM_{2}$ we have to fix some model parameters. Here, we adopt a standard paramagnetic {\it s-p} density of
states extracted from first-principles calculations. The exchange splitting in the {\it s-p} energy bands induced by the local moments of the rare-earth ions, was properly chosen to yield the {\it s-p} magnetic moment at the $R$ sites of the host, which is assumed to be of the order of 0.1 of the {\it d} magnetization at the $R$ sites. The parameter $\alpha $ which renormalizes the hopping energy, was chosen $\alpha\simeq 1$ giving the ratio between the extension of the host and impurity {\it s-p} wave functions. For the whole rare earth series we adopted $J^{sd} = 0.2 \times 10^{-3}$ in units of the {\it s-p} bandwidth in the case of $R$Fe$_{2}$, while for $R$Co$_{2}$, we adopted $J^{sd} = 0.4 \times 10^{-3}$ for heavy rare-earth ions and $J^{sd}=0.5 \times 10^{-3}$ for light rare-earth ions both also in units of the {\it s-p} bandwidth. These choices are enough to reproduce the systematics of the respective series. The magnetic moments at a Fe site in $R$Fe$_{2}$ $\left( \left\langle S^{{\rm Fe}}\right\rangle \right) $ and at a Co site in $RCo_{2}$ $\left( \left\langle S^{{\rm Co}}\right\rangle \right) $ were obtained from Ref.~\cite{Delyagin2007}. In the present case, since we are discussing Laves phase compounds, $Z_{{\rm nn}}=12$. Keeping fixed these parameters, we self-consistently determined the local magnetic moment and the corresponding magnetic hyperfine field at the Cd or Sn impurity. 
Fig.~\ref{fig:dos} exhibits a typical local density of states (a) at the Cd impurity, and (b) at the  Sn impurity diluted at $R$Co$_{2}$, namely for $R$ = Gd (Cd is not soluble\cite{priva} in $R$Fe$_{2}$). The local charge potential $V_{0}^{\sigma }$ and the exchange interaction with the $M$ ions are such as to produce a slight deformation and a shift of the up and down sub-bands generating a total local magnetic moment in down direction for the Cd impurity 
and up direction for Sn impurity (for the latter see also Fig.~\ref{fig:mmSn}~(b)). 
In fact the Cd impurity local DOS- here $\Delta Z =1.2$- is close to the rare earth {\it s-p} DOS, which has also a negative {\it s-p} magnetization. On the other hand for the Sn impurity the local potential- now $\Delta Z =3.2$ -, piles up the down and the up spin states in such way that the contribution of the up states to the DOS overcomes that of the down, thus generating a change of sign of the magnetic moment with respect to the one from the {\it s-p} rare earth. 
In  table \ref{tab:SnRFe2} and \ref{tab:SnRCo2}, the calculated contribution to the magnetic hyperfine fields as well the experimental measurements collected from Ref.~\cite{Delyagin2007} for a Sn impurity diluted at $R$ site of the $R$Fe$_{2}$and $R$Co$_{2}$ respectively are displayed. The results of a Cd impurity at $R$Co$_{2}$ are shown in the table \ref{tab:bhfcd}. In both tables the agreement with the experimental data are fairly good.
\begin{table}[h]
\caption{Calculated contributions (from $R$, induced and total) to the magnetic hyperfine fields, in Tesla, at Sn impurity in $R$Fe$_{2}$. Experimental data were collected from Ref.~\cite{Delyagin2007}. The mean-square errors is about $0.6$~T.}
\begin{tabular}{llccccc}
\hline\hline
             &    $R$           & Gd         & Tb         & Dy          & Ho          & Er          \\ \hline 
             &$B_{hf}^{R}$      & $-$4.5     & $-$3.4     & $-$2.6      & $-$2.3      & $-$2.1        \\  
	     &$B_{hf}^{\rm ind}$& 56.9       & 53.7       & 52.5        & 49.6        & 49.3        \\    
             &$B_{hf}$          & 52.4       & 50.3       & 49.9        & 47.3        & 47.2        \\ \hline  
Exp.         &$B_{hf}$          & 52.6       & 51.2       & 48.8        & 47.5        & 48.0        \\ \hline\hline
\end{tabular}
\label{tab:SnRFe2}
\end{table}
\begin{table}[h]
\caption{Calculated contributions (from $R$, induced and total) to the magnetic hyperfine fields, in Tesla, at Sn impurity in $R$Co$_{2}$. Experimental data were collected from Ref.~\cite{Delyagin2007}. The mean-square errors is about $0.6$~T.}
\begin{tabular}{llccccc}
\hline\hline
             &    $R$           & Gd         & Tb         & Dy          & Ho          & Er        \\ \hline 
             &$B_{hf}^{R}$      & $-$5.3     & $-$2.1     & $-$0.9      & $-$0.4      & $-$0.4        \\  
             &$B_{hf}^{\rm ind}$& 57.2       & 50.9       & 45.7        & 43.9        & 42.7        \\  
             &$B_{hf}$          & 51.9       & 48.8       & 44.8        & 43.5        & 42.3        \\ \hline
Exp.         &$B_{hf}$\,      & 52.1         & 47.9       & 45.0        & 43.4        & 41.3        \\ \hline\hline
\end{tabular}
\label{tab:SnRCo2}
\end{table}
\begin{table}[h]
\small{
\caption{Calculated contributions (from $R$, induced and total) to the magnetic hyperfine fields, in Tesla, at Cd impurity in $R$Co$_{2}$. Experimental data were collected from Ref.~\cite{Presa2000}. The mean-square errors is about $0.09$~T.}
\begin{tabular}{llccccccccc}
\hline\hline

& $R$               &Pr      &Nd      &Pm      &Sm     &  Gd         & Tb           & Dy           &  Ho          & Er    \\ \hline
& $B_{hf}^{R}$      &$-$5.43 &$-$5.00 &$-$3.50 &$-$2.00&$-$7.57  & $-$6.36      &  $-$5.66     &  $-$5.03     & $-$4.62 \\  
& $B_{hf}^{\rm ind}$&$-$7.57 &$-$10.95&$-$13.55&$-$16.14&$-$13.63& $-$13.20     & $-$12.74     & $-$12.68     & $-$12.56 \\  
& $B_{hf}$          &$-$13.00&$-$15.95&$-$17.05&$-$18.14&$-$21.20& $-$19.56     & $-$18.40     & $-$17.71     & $-$17.18 \\ \hline 
Exp.& $|B_{hf}|$    &12.61   &15.98   & ---    &18.17   &  21.18 & 19.49        & 18.30        & 17.66        & 17.22   \\ 

\hline\hline 
\end{tabular}
\label{tab:bhfcd}
}
\end{table}

\begin{figure}[htbp]
\includegraphics[angle=0,width=0.60\textwidth]{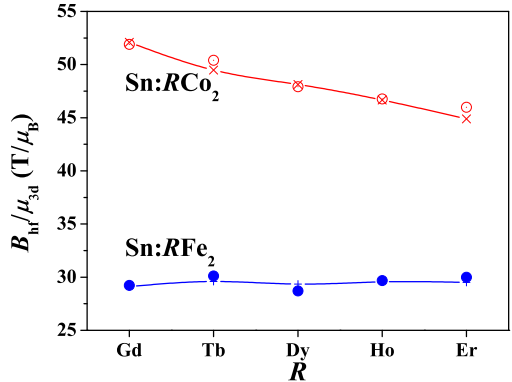}
\caption{Calculated ratio $B_\mathrm{hf}/\mu_{3d}$ at Sn impurity in $R$Co$_{2}$ (full line) and $R$Fe$_{2}$ (dashed line). The open circles and full circles represent their respective experimental data displayed in tables \ref{tab:SnRFe2} and \ref{tab:SnRCo2}. The ratio is almost constant in $R$Fe$_{2}$ but varies in $R$Co$_{2}$ along the rare earth series (see text).} 
\label{fig:bhfsn}
\end{figure}%

Figure \ref{fig:mmSn}~shows the self-consistently calculated magnetic moments at a Sn impurity site in (a) $R$Fe$_{2}$ and (b) $R$Co$_{2}$. From this figure one can see that the contribution to the local magnetic moment arising from the 3d neighboring ions, $\widetilde{m}^{\rm ind}_{0}$, is larger than the one originated from the rare-earth ions $ \widetilde{m}^{R}_{0}$, in modulus. 

\begin{figure}[htbp]
\includegraphics[angle=0,width=0.60\textwidth]{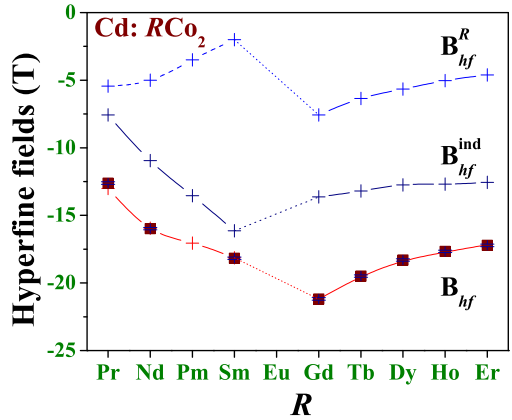}
\caption{Calculated contributions to the total magnetic hyperfine fields $B_{\rm hf}$ at Cd impurity in $R$Co$_{2}$ (solid line). The dotted line correspond to the contribution from the rare earth and the dashed from the Co ions, both  with the same sign. The squares represent experimental data, based on table~\ref{tab:bhfcd}.} 
\label{fig:bhfcd}
\end{figure}

\begin{figure}[htbp]
\includegraphics[angle=0,width=0.60\textwidth]{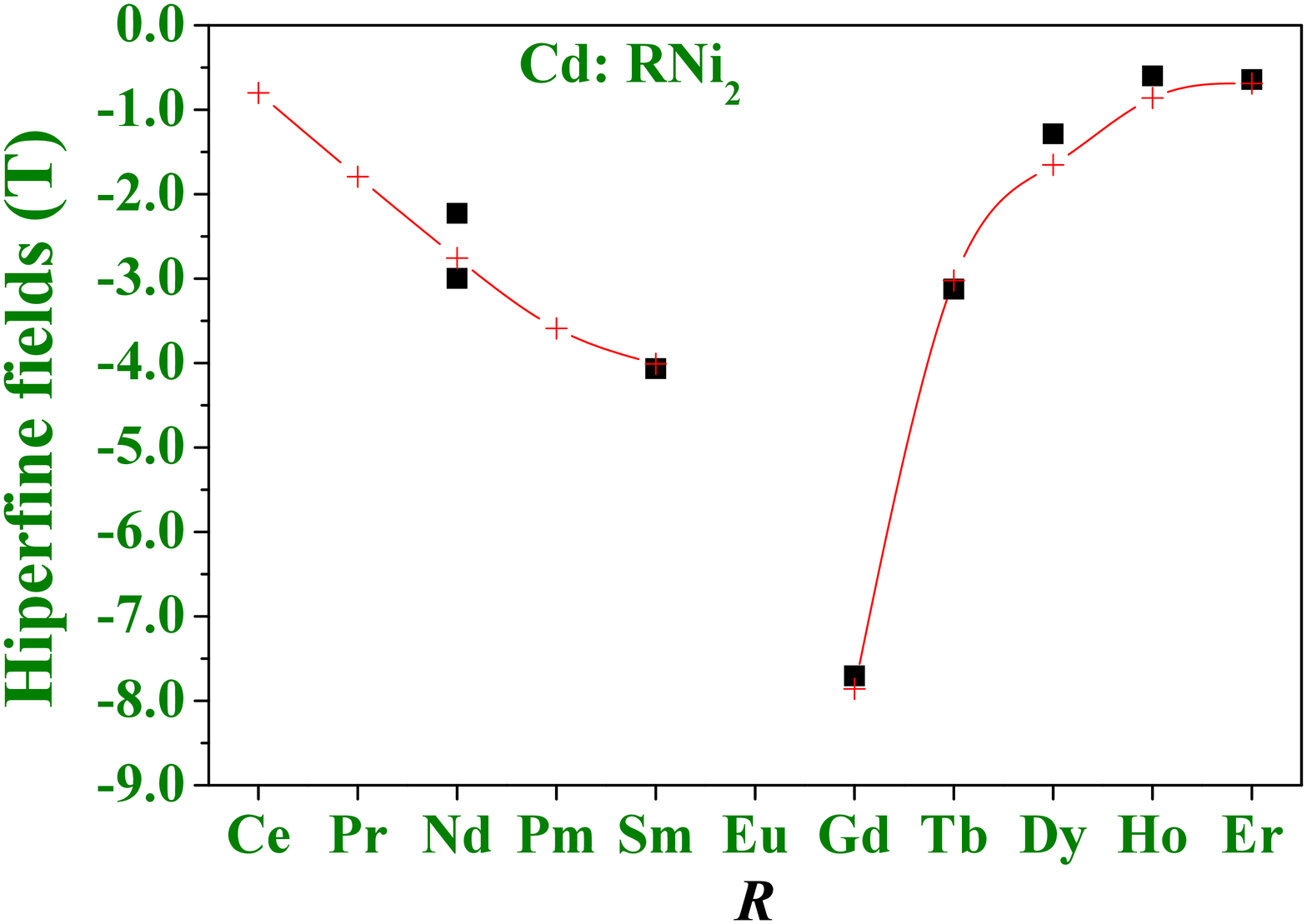}
\caption{Calculated magnetic hyperfine fields at Cd impurity diluted in $R$Ni$_{2}$. Squares represent experimental data.  Now the hyperfine field varies along the rare earth series. See text.} \label{fig:cdrni2}
\end{figure}
Figure \ref{fig:bhfsn} exhibits the $B_\mathrm{hf}/\mu_{3d}$ ratio, in T/$\mu_{B}$, where $\mu_{3d}$ is the magnetic moment of the $3d$ (Fe or Co) atom.  We can observe the different behavior between Sn:$R$Co$_{2}$ and Sn:$R$Fe$_{2}$ systematics.  While in the $R$Fe$_{2}$ series the $B_\mathrm{hf}/\mu_{3d}$ ratio remains almost constant, in  $R$Co$_{2}$ the ratio decreases from $R$ = Gd to $R$ = Er.
In both compounds the dominant contribution is from the $3d$ ions (see tables~\ref{tab:SnRFe2} and \ref{tab:SnRCo2}), the contribution from the rare-earth ions being very small. However, the amplitude of the total hyperfine field is about twice larger in $R$Co$_{2}$ than in $R$Fe$_{2}$ as one goes from Gd to Er.  In addition, Co has  $\mu_{3d} \approx 1\mu_{B}$ along the series while Fe has about twice larger $\mu_{3d}$. The combination of these aspects, and not an absent contribution from the rare-earth sublattice~\cite{Delyagin2007}, reduce the ratio amplitude in $R$Fe$_{2}$ by a factor $\approx 4$.

Fig.~\ref{fig:bhfcd} exhibits the contributions to the total magnetic hyperfine fields at the Cd impurity diluted in $R$Co$_{2}$. In this case, $B_{hf}^{R}$ and $B_{hf}^{\rm ind}$ have the same sign in agreement with the previous discussion  about figure~\ref{fig:dos}.

Now, a few remarks concerning the magnetic hyperfine field at Cd impurity in $R$Ni$_{2}$, where some experimental data are available~\cite{Presa2004}. In this case, the d-band associated to the Ni sublattice is completely filled so no induced magnetic moment produced by Ni ions at the Cd impurity is present. Then, as illustrated in Figure \ref{fig:cdrni2}, only contribution due to the rare earth sublattice is present. The calculated $B_\mathrm{hf}$ is thus smaller, in absolute value, than those of $R$Co$_{2}$.

\section*{Acknowledgments}
We acknowledge the support from the Brazilian agencies PCI/MCT and CNPq and useful discussion with Prof. M. Forker and Prof. H. Saitovich.


\begin{thebibliography}{00}

\bibitem{Oliveira2003}A. L. de Oliveira, N. A. de Oliveira and A Troper, Physical Review{\bf B 67}, 012411, 2003.

\bibitem{Acker91b} J.F. van Acker , W. Speier, J.C. Fuggle and R. Zeller, Physical Review{\bf  B 43}, 13916, 1991.

\bibitem{Oliveira95}N. A. de Oliveira, A. A. Gomes and A. Troper, Phys. Rev. {\bf B52}, 9137, 1995. 

\bibitem{Delyagin2007}N. N. Delyagin and V. I. Krylov, J. Phys.: Condens. Matter {\bf 19}, 086205, 2007.
\bibitem{priva} M. Forker, private communication.

\bibitem{Presa2000} P. de la Presa , S. M\"{u}ller, A. F. Pasquevich and M. Forker, J. Phys.: Condens. Matter~{\bf 12}, 3423, 2000.

\bibitem{Presa2004} S. M\"{u}ller, P. de la Presa and M. Forker, Hyperfine Interact.~{\bf 158}, 163, 2004.


\end{thebibliography}
\end{document}